# Some *A Priori* Torah Decryption Principles[1]

**Grenville J. Croll**

grenville@croll-management.freeserve.co.uk

**The author proposes, *a priori*, a simple set of principles that can be developed into a range of algorithms by which means the Torah might be decoded. It is assumed that the Torah is some form of transposition cipher with the unusual property that the plain text of the Torah may also be the cipher text of one or more other documents written in Biblical Hebrew. The decryption principles are based upon the use of Equidistant Letter Sequences (ELS's) and the notions of Message Length, Dimensionality, Euclidean Dimension, Topology, Read Direction, Skip Distance and Offset. The principles can be applied recursively and define numerous large subsets of the 304,807! theoretically possible permutations of the characters of the Torah.**

## 1. Introduction

As a result of the controversy generated by the "Bible Code" book by Michael Drosnin [8] and the resultant publicity given to the work of Witztum, Rips & Rosenberg [1] (WRR), the author was motivated to develop this set of a priori Torah decryption principles in late 1997. They are described with some additional commentary and provide the background for other more recent work completed by the author.

Controversy aside, the author maintains a working hypothesis that the Torah is a cryptogram and that it is intended that it be decoded. More specifically, the author's view is that the Torah is some form of transposition cipher[2] and that ELS's are the basis for any possible decryption. The author takes the view that the Torah may have an unusual property in that the plain text of the Torah may also be the cipher text for at least one other lengthy document written in Biblical Hebrew.

Though the above suggestion may seem ludicrous, the existence of anagrams is unremarkable and exemplify the principle of readable text being readily concealed within other readable text. Of course, the length of a traditional anagram is very short in comparison with the proposition being made with regard to the Torah. Long transposition ciphers with a lucid cipher text are exceedingly difficult to create [9] using conventional computation and it is possible that this difficulty may have been overcome in respect of the Torah by the use of some form of Quantum Computer[3].

Note that despite the extreme difficulty of creation of long transposition ciphers (using conventional finite state automata), it is entirely possible that decryption (using conventional finite state automata) may be relatively easy. In this respect there may turn out to be a direct analogy between the RSA cipher[4] and the Torah. RSA uses the relative difficulty of factorisation of large integers as its Trap Door function. Likewise, the Torah may have some transposition algorithm and a long key as the basis for its own one way barrier function.

The author is aware of the status of the Torah and ignores the obvious theological and philosophical problems that work in this area poses, save to note that should it turn out to be the case that the Torah is a large scale (possibly multiple) transposition cipher, then there may be severe limitations as to speculation with regard to its origin.

Since the method of any encoding and decoding of the Torah are unknown to us, perhaps we should approach the matter in some sort of logical sequence.

---

# 2. Use of ELS's

## 2.1 Background

Much has been written regarding ELS based codes in the Torah and other secular and religious material in a wide variety of languages. Though attempts have been made to establish rigorous proofs that there are or are not [11] ELS based "codes" in the Torah, the issue is not resolved. For the proponents there remains a series of stunning coincidences which defy statistical analysis due to their uniqueness. For the sceptic, the Torah remains an ordinary document that appears entirely random when subjected to a variety of statistical analyses [10] [12]. The author has results that fit squarely in both camps.

Common factors in most forms of this work are the use of ELS's, per WRR's original definition, though some unreported work has been done on Gradually Increasing/Decreasing Letter Sequences. A further and more imposing common factor is that most work so far has focussed upon the use of ELS's and the attempted detection of proximate conceptually related word pairs. Few attempts have been made to widen the field to embrace other potential properties of the Torah and other textual material.

In this paper, though ELS's are used as the basis for the various algorithms, there is almost no commonality with other previous work in this area.

## 2.1 Algorithm One

It is apparent that if we skip through texts whose length L is prime, using a skip distance D where $1 < D < L$, we can transpose an original text into L-2 different derivative texts very easily:

Let T[*i*] be the *i*th character of an array containing the Torah, or some other section of the Hebrew Bible or a control and P[*i*] be the *i*th character of an array containing the permuted document. If the lengths of T and P are L and L is prime and D is the ELS skip distance, then we can permute the input document using the simple 'C' routine:

### /* Algorithm One */

```
for (i=0;i<L;i++)
 {
 P [i] = T [D*i % L];  /* % is the 'C' modulus operator */
 };
```

For example, the text string "MARY HAD A LITTLE LAMB!" which is 23 characters long including spaces, can be transposed into the nonsensical string "MR A ITELM!AYHDALTL AB" by considering it as a one dimensional ring of characters and using a skip distance (or "key") of two.

There are 23 factorial permutations of a meaningful 23 character string and it is extremely unlikely that any of the 21 permutations defined above should result in a readable derivative text. For longer strings, the probability of finding a meaningful sequence in one of the derivative permutations tends to zero at the super exponential rate of 1/(N-1)!

Though vanishingly small, the possibility cannot be excluded that there exists an algorithm (perhaps based upon Algorithm One) which might permute the characters of the Torah into some other document as a result of wilful and very clever design.

By way of further example, Table 1 contains a list of English words that can be permuted into other English words using Algorithm One and a skip distance of two. Table 1 was compiled by manually searching the pages of a pocket dictionary. Readers might like to attempt the creation of longer, meaningful word strings such that they permute into other meaningful word strings by the simple application of Algorithm One.

## Table 1

| | |
|---|---|
| Feast | Fates |
| Fount | Futon |
| Green | Genre |
| Point | Piton |
| Treat | Tetra |
| Moans | Mason |
| Pearl | Paler |
| Perry | Pryer |
| Prise | Piers |
| Taint | Titan |
| Weird | Wider |

Note that Algorithm One significantly reduces the number of anagram candidates. Many five character words are anagrammatical. However, of the 120 possible permutations of the characters of a five character word, Algorithm One only defines five of them. This is important when considering longer strings and more complex algorithms because the number of possible permutations becomes effectively infinite (in fact uncountably finite). Algorithm One, its derivations and variations address the computable[5] permutations of the characters of the Torah.

# 4. Establishing Message Length

## 4.1 Data Communications fundamentals

One of the fundamental issues in the field of Data Communications is the means by which the length of a transmitted section of text is communicated between sender and recipient. Techniques include the use of blocks of a pre-determined size and variable length blocks with a pre-specified delimiter. It is not unreasonable to assume that message texts from unknown sources might have a length which was prime. Note that the 3 Terawatt SETI message transmitted from the Arecibo radio telescope in 1974 consisted of 1679 bits in a 73 * 23 B&W binary image.

Other potentially relevant data communications techniques include the use of error detection and correction methods and the use of data compression techniques to reduce the data volume and transmission time. Basic techniques include parity checking, Cyclic Redundancy Checks, Hamming codes and so on. Which, if any, of these techniques might be of relevance in analysing the Torah remains to be seen.

In the development of the algorithms that follow, no assumption is made about the data that is being analysed. Once a text and a message length has been chosen, the various approaches to further processing it naturally follow.

## 4.1 Torah Message Length

There are numerous versions of the Torah, whose content and length vary by a small amount.

The traditions surrounding the historic copying and distribution of the Torah are strict and have led to the near letter perfect preservation of the various versions of the Torah for in excess of a thousand years. WRR and the author base their work upon the digital version of the Koren edition[6] of the Torah. The Koren edition is itself based upon the ancient Masoretic Textus Receptus. The author assumes that the particular features of the Koren edition of the Torah are there by design and various inferences are drawn. There is no reason however why differing inferences should not be drawn or that these methods should not be applied to any other version of the Torah (eg the Leningrad Codex), or any other document in whatever language.

The Koren edition of the Torah is 304,807 characters long including two inverted letter nuns.

304,807 is prime.

### 4.2 The Inverted Letter Nuns

There are 22 basic characters in the Hebrew alphabet and in addition there are two instances where the letter nun, which is the 14th character in the Hebrew alphabet, is written upside down in the Torah. The inverted letter nuns bracket the short section of text comprising Numbers 10 verses 35 & 36:

35 When the Ark would journey, Moses said "Arise HASHEM, and let
Your foes be scattered, let those who hate You flee from before you"
36 And when it rested, he would say, "Reside tranquilly, O HASHEM,
among the myriad thousands of Israel". [7]

Nobody knows why the inverted letter nuns are there. One theory has it that the short section of text between the inverted nuns is a book in its own right.

Due to the unusual nature of the inverted nuns (especially their resemblance to a pair of ASCII square brackets), we make an assumption that their presence is intended to draw attention to the fact that there are three distinct textual sections within the Torah. We assume that their presence and absence, together with the presence and absence of the section of text between the inverted nuns and the inverted nuns themselves is related to the issue of establishing message length.

Table 2 sets out some of the possible combinations of text sections and gives the prime factors of the lengths of each of the various combinations of the sections of text. We refer to the three major text sections as T1, T2 and T3 and the letter nuns as N1 and N2.

By inspection, the length of the text between the inverted letter nuns in the Koren edition is 85 characters, thus text section T2 has length 85. The lengths of sections T1 and T3 are 206,588 and 98,132 respectively.

Table 2

| Text | Length | Number of Factors | Prime Factors |
|---|---|---|---|
| T1,N1,T2,N2,T3 | 304807 | 1 | 304807 |
| T1,T2,T3 | 304805 | 2 | 5, 60961 |
| T2 | 85 | 2 | 5, 17 |
| T1 | 206588 | 3 | 2, 2, 51647 |
| T3 | 98132 | 3 | 2, 2, 24533 |

Other combinations of text section can be examined (including or not including inverted nuns)

## 5. Dimensionality & Euclidean Dimension

Given the factorisation characteristics of the various text sections as given in Table 2 we could consider the resultant text as either a linear string, a two dimensional array, a three dimensional cuboid or some other shape in M dimensions.

Clearly, with a regular, M-dimensional layout of text, the process of skipping through the text (along whichever dimension) to generate further derivative texts using variations of Algorithm One is made much simpler when there are no jagged edges.

Note that WRR's analysis assumes that the text of the book of Genesis is laid out in two dimensions, in rectangular arrays of varying size with an incomplete last row. The irregularity of these shapes would not permit the efficient creation of large derivative texts using ELS's and a simple algorithm.

# 6. Linear permutation

For a text of 304,807 characters there are 304,805 permutations of this text using ELS's and Algorithm One as described in section 3. Automatic evaluation of these 304,805 derivative versions of the Torah using software designed to either detect meaningful Hebrew text or measure randomness is not difficult.

Examination of a string of length L where L is prime using Algorithm One is referred to a Linear Permutation. Use of **Algorithm Two** (the Maxquad function) to measure the randomness of the Linear Permutations is reported separately.

# 7. Rectangular permutation

## 7.1 Methods of arranging text

If we now turn our attention to the second row of Table 2 where T1, T2 & T3 are concatenated, we note that the length of the document is 304,805, which factorises into 5 * 60,961, potentially implying that the text should be laid out in a two dimensional array comprising 5 rows of 60961 columns.

We have no information however about how we should arrange this text within the confines of a two dimensional array. Is the text to be treated as a five long strings, each of length 60961, or is it to be treated as 60961 strings each of length 5?. Do we arrange the text in each row right to left or left to right? Is the text to be treated as a 5 * 60961 ring, or is it to be treated as a 5 * 60961 Mobius ring? Or is the topology more complicated still?

## 7.2 Textual Topology & Read Direction

Thus the establishment of the Topology and text Read Direction of any encoding in two, three and more dimensions is crucial if we are to fully investigate permutations of the Torah.

In considering the topology as being either a ring or a Mobius ring, we note that these are two specific variations of the 5! different ways of joining up the ends of a 5 * 60961 array. In fact there are 120 different topologies obtained by placing the five rows in each of 120 different combinations.

Of course, we do not know if we are to arrange the text left-to-right or right-to-left within each row. Since there are 5 rows, there are 2^5 or 32 different ways we can arrange the rows, and each of these should be considered.

## 7.3 Skip Distance

Having arranged the text in rows and columns in two dimensions using all the various combinations of topology and read direction, we can then use a simple variation of Algorithm One to skip through the complete two dimensional array every D characters, starting at the first character and omitting values for D which are not co-prime to 5 and 60961 (or to each of the vertex lengths if the dimensionality is not two).

## 7.4 Offset

Of course, starting Algorithm One at the first character is restrictive - there is no reason why Algorithm One could not be started at some other character position - the Offset position. The same character sequence is generated, but the resultant string is byte shifted by the Offset amount. Offset only becomes significant when considering recursion as otherwise the same sequences are generated, rotationally shifted by one or more bytes.

## 7.5 Tractability

Note that in the above discussion, we assume that the text is being laid out in 5 long strings each of length 60961 which gives rise to the possibility of there being 5! differing topologies and 2^5 different row read directions. Of course, we could just as easily assume that there are 60961 strings of length 5 to be placed in 60961! different topologies with 2^60961 row read directions. Clearly, the former is tractable, whereas the latter is not (yet). For this reason, and in all further analysis, text is laid out in a smaller number of longer rows and ELS skipping takes place along the long direction only.

## 7.6 Recursion

Examination of a string of length L where L has two prime factors using the method outlined above is referred to as Rectangular Permutation. The resultant algorithm is known as **Algorithm Three**. Algorithm Three is recursive and recursion can take place to any number of Levels, though at higher levels of recursion the issue of computability becomes relevant. Algorithm Three is specified more exactly in Appendix A.

## 7.7 Number of Permutations

For Torah text T1T2T3 and T2 alone, the total number of permutations of the text using Algorithm Three to various levels of recursion is:

| | T1T2T3 | T2 |
|---|---|---|
| Topologies | 120 | 120 |
| Read Directions | 32 | 32 |
| ELS Skips | 243839 | 32 |
| Offsets | 304805 | 85 |
| | | |
| Recursion Level 1 | 2.85E+14 | 1.04E+07 |
| Recursion Level 2 | 8.15E+28 | 1.09E+14 |
| Recursion Level 3 | 2.32E+43 | 1.14E+21 |
| Recursion Level 4 | 6.63E+57 | 1.19E+28 |
| Recursion Level 5 | 1.89E+72 | 1.24E+35 |

Note that T1T2T3 and T2 are similar in that they both have 5 rows. There is therefore a possibility that T2 may be the index for the whole Torah. Any Algorithm Three permutations of T2 that are lucid may be indicative of the possibility that application of the same Algorithm Three parameter sequence to T1T2T3 may result in the derivation of a New Torah.

Based upon estimates of currently available technology a single large device could be built to achieve exhaustive evaluation of T2 to Recursion Level 3.

## 8 Some A priori short cuts

Given the substantial number of permutations of even T2 that can be evaluated, there exists the possibility that any encoder may have built in some clues as to the correct topology and read direction in the design of the message itself. This might avoid exhaustive evaluation of every possible combination.

## 8. 1 The Topological Interlock

The author proposes (*a priori*) that if the text of the Torah is laid out as a 5*60961 array (or a 5*17 array as in the case of T2) and is examined using ELS sequences in the long and short directions with the same key, for one topology (being the correct topology) the two sub components generated during this process may be the same, similar or complementary. The author denotes this test as the Topological Interlock test. The two components being the horizontal and vertical Topological Interlock components.

For example, assume a semi-nonsense 21 character plain text "GOSSIERMISNOMEREXODUS" laid out on a 7 by 3 array. The rows are numbered 1 to 3 and the columns labelled A to G, in standard spreadsheet format. Using a skip of 5, the Horizontal and Vertical Topological Interlock Components are calculated as follows:

```
A B C D E F G                Vertical Component

1     G O S S I E R          A1 B3 D2 F1 G3 B2 D1
2     M I S N O M E          E3 G2 B1 C3 E2 G1 A3
3     R E X O D U S          C2 E1 F3 A2 C1 D3 F2

Horizontal Component

A1 F1 D2 B3 G3 E1 C2
A2 F3 D1 B2 G2 E3 C1
A2 F2 D3 B1 G1 E2 C3
```

The actual text of the two interlock components is therefore:

```
      A B C D E F G          Vertical Component

1     G O S S I E R          G E N E S I S
2     M I S N O M E          D E O X O R R
3     R E X O D U S          S I U M S O M

      Horizontal Component

      G E N E S I S
      R U S I E D S
      M M O O R O X
```

One row of the interlock components match exactly and indeed spell an interesting word "GENESIS". We may therefore be drawn to the conclusion that the 7 by 3 layout with traditional English left-right top-bottom reading is the correct topology for further investigation of this text.

## 8. 2 The Directional Interlock

Similarly, in considering text laid out in a 5 * 60961 array in what may have been determined to be the correct topology, there are four directions in which text can then be read starting from the first character.

These are denoted North, South, East and West. The author proposes that read direction may also be indicated by a Directional Interlock test, where the correct read direction is indicated by one of the four Directional Interlock components being distinctly different from the other three, when each component is created using the same ELS key.

Let us again assume that we are examining some text again laid out on a 7 * 3 array. Using spreadsheet notation and a key of 5, we can derive the North, South, East and West components of the Directional Interlock.

```
                                North

                                A1 C2 E3 F1 A2 C3 D1
                                F2 A3 B1 D2 F3 G1 B2
                                D3 E1 G2 B3 C1 E2 G3

     West                       Torah                   East

                                  A B C D E F G
A1 C2 E3 G1 B1 D2 F3          1 T   E O L E S        A1 F1 D2 B3 G3 E1 C2
A3 C1 E2 G3 B3 D1 F2          2   A H D C N E        A3 F3 D1 B2 G2 E3 C1
A2 C3 E1 G2 B2 D3 F1          3 R I P S I I T       A2 F2 D3 B1 G1 E2 C3

                                     South

                                A1 B3 D2 F1 G3 B2 D1
                                E3 G2 B1 C3 E2 G1 A3
                                C2 E1 F3 A2 C1 D3 F2
```

The texts of the four directional interlock components are therefore:.

```
                                   North

                                T H I E   P O
                                 N R   D I S A
                                S L E I E C T

          West                       Torah                East

     T H I S   D I              T   E O L E S         T E D I T L H
     R E C T I O N             A H D C N E            R I O A E I E
       P L E A S E              R I P S I I T         N S   S C P

                                   South

                                T I D E T A O
                                I E   P C S R
                                H L I   E S N
```

In the above example there is ample evidence to suggest that West is the correct read direction.

## 8. Cubic Permutation

T1 and T3 each have three prime factors and can be permuted by algorithms similar to those above, with the additional factor being that there are eight choices of corner from where the ELS skip through could commence from.

## 9. Summary & Conclusion

Having given due consideration to the establishment of Message Length, Dimensionality, Euclidean Dimension, Topology, Read Direction, Skip Distance and Offset, the Torah and various parts of it can be recursively permuted into myriad further long strings using a simple algorithm. The order in which these various steps can be applied to the Torah can be varied in a relatively limited way.

Given the extreme difficulty of creating long transposition ciphers where the cipher text is also lucid, it is incredibly unlikely that even the short section of text comprising T2 should be capable of permutation into a readable sequence by any variation of the algorithms described above (or any other algorithms), except by virtue of intentional design.

The task of evaluating the above algorithms and any output will be very substantial. However, Algorithm Three is suitable for execution on highly parallel architectures and also in specialist PLA type hardware where sequence generation (and some evaluation) can take place at clock speed rates.

To exhaustively evaluate T2 to recursion Level 3 will require, for example, the development of specialist hardware in substantial volume operating within a large networked configuration. Alternatively, if sufficient publicity can be generated, a significant proportion of the machines connected to the internet operating in unison could be just as effective.

If it is thought that T2 may be the index for T1T2T3, then the parameters from T2 sequences that pass a priori readability or lucidity thresholds can be applied automatically to T1T2T3.

## Acknowledgements

The author gratefully acknowledges the support and assistance provided by the staff of Eastern Software Publishing Ltd (ESP) during the course of this project. This paper is dedicated to the memory of Julia Bryans of ESP, whose diligent efforts over several years helped to financially support this work.

## References

[1] Witztum, Rips & Rosenberg, Equidistant Letter Sequences in the Book of Genesis, Statistical Science, 1994, Volume 9, No 3, 429-438

[2] Schneier, B. Applied Cryptography, 2nd Edition, (1996), John Wiley & Sons

[3] Deutsch, D. "Quantum Theory, the Church-Turing Principle, and the Universal Quantum Computer" Proceedings of the Royal Society, London, Vol A400, (1985), pp 97-117.

[4] Rivest, R., Shamir, A., and Adelman, L. "A Method of Obtaining Digital Signatures and Public Key Cryptosystems", Communications of the ACM, Vol 21 (1978) pp 120-126

[5] Turing, A.M. "On Computable Numbers with an Application to the Entsheidungsproblem", Proceedings of the London Mathematical Society, Vol 42, (1937)

[6] Bible Codes, Computronic Inc, (1997) P.O. Box 102, Savyon, Israel.

[7] Nosson Scherman (Ed), Tanach (Hebrew Bible), Stone Edition, Mesorah Publications Ltd. NY

[8] Drosnin, Michael "The Bible Code", Simon & Schuster, New York, 1997

[9] Ritter, T. "Transposition Cipher with Pseudo-Random Shuffling: The Dynamic Transposition Combiner", Cryptologia, 1991, Vol 15 Number 1 pp1-17

[10] McKay, B. "Solving the Bible Code", Statistical Science, 2000

[11] Nichols, R.K. "The Bible Code", Cryptologia, 1998 Vol 22 Number 2 pp 121-133

[12] Ingermanson R. S. "Who wrote the Bible Code", Waterbrook Press, Colorado Springs, 1999

## Appendix A  Algorithm Three

Here is some indicative 'C' code for Algorithm Three:

```
int t2[85], loop_t2[85], rp_t2[85];                    // T2, TEMP T2, PERMUTED T2

int topology [120] [5] = {                             // TOPOLOGY TABLE
                         { 0, 1, 2, 3, 4},
                         { 0, 1, 2, 4, 3},
                            etc                        // 120 Rows Long
                         { 4, 3, 2, 0, 1},
                         { 4, 3, 2, 1, 0},
                         };

int top, row, offset, skip;
int r0, r1, r2, r3, r4;                                // Row Flip Binary Counters
int i, j, k, m, u;                                     // Loop Counters etc

for (top=0;top<120;top++)                              // 120 Topologies
{ i=0;
  for (u=0;u<5;u++)                                    // Load the 5 rows in order
 {for(m=0;m<17;m++)                                    // Each of 17 Characters
   {loop_t2[i++]=t2[(17*topology[top][u])+m];};        // Depending upon which Topology
  };
 for(r0=0;r0<2;r0++)
 { reverse_row(loop_t2,0);                             // Left Right Reverse The First Row
   for(r1=0;r1<2;r1++)
   { reverse_row(loop_t2,1);                           // Left Right Reverse The Second Row
     for(r2=0;r2<2;r2++)
     { reverse_row(loop_t2,2);                         // Left Right Reverse The Third Row
      for(r3=0;r3<2;r3++)
      {reverse_row(loop_t2,3);                         // Left Right Reverse The Fourth Row
       for(r4=0;r4<2;r4++)
       {reverse_row(loop_t2,4);                        // Left Right Reverse The Fifth Row
        for(offset=0;offset<85;offset++)               // For 85 Offsets
         {for(skip=0;skip<85;skip++)
          {if(skip % 5 !=0 && skip % 17 !=0)           // For (net) 32 Skips
           {for(m=0;m<85;m++)
                 {rp_t2[m]=loop_t2[(m*skip+offset)%85];}       // Permute T2
            analyze(rp_t2);                            // Analyse or pass to next Recursion Level
            };
          };
         };
        };
       };
      };
     };
    };
   };
                                                       // Don't expect this to compile!
```